**This is the accepted manuscript (postprint) of the following article:**
M. Shaker, E. Salahinejad, W. Cao, X. Meng, V.Z. Asl, Q. Ge, *The effect of graphene orientation on permeability and corrosion initiation under composite coatings*, Construction and Building Materials, 319 (2022) 126080.
https://doi.org/10.1016/j.conbuildmat.2021.126080

# The effect of graphene orientation on permeability and corrosion initiation under composite coatings

Majid Shaker [a], Erfan Salahinejad [b,*], Weiqi Cao [a,c], Xiaomin Meng [a], Vahdat Zahedi Asl [d], Qi Ge [a]

[a] Chongqing 2D Materials Institute, Liangjiang New Area, Chongqing, 400714, China

[b] Faculty of Materials Science and Engineering, K. N. Toosi University of Technology, Tehran, Iran

[c] Key Laboratory of Optoelectronic Technology & System Ministry of Education, Chongqing University, Chongqing, 40044, China.

[d] College of Materials Science and Engineering, Beijing University of Chemical Technology, Beijing, China

*Corresponding Author:

E-mail address: salahinejad@kntu.ac.ir (Erfan Salahinejad)

## Abstract

The unique anisotropic properties of graphene, particularly impermeability, have made it a promising candidate for further advances in corrosion prevention applications. Despite the large number of experimental works divulging the use of graphene in anticorrosion coatings, there is no report on the numerical modelling and simulation of the relationships between the orientation of graphene sheets in composite coatings and the introduced corrosion protection efficiency, to our knowledge. Herein, it is tried to model the influence of the orientation of graphene sheets dispersed in organic coatings on the diffusivity and flux of corrosive substances besides the corrosion initiation time of the protected substrates. To discover the relationship between the graphene orientation and corrosion-related phenomena, this study introduces a novel model consisting of a trigonometric factor named unprotected projected surface area proportion, which calculates corrosion-related parameters based on the principal Fick's laws. The model reveals that the decrease in the angle between graphene sheets and the substrate is highly beneficial for postponing the corrosion onset. It is accordingly estimated that a mismatch angle of 10º can slow down the diffusion process significantly and delay the corrosion initiation by around 65 times in a 100 µm thick epoxy/graphene composite coating in comparison to the counterpart with the perpendicular alignment. The predicted corrosion parameters were in a good agreement with the experimental data, indicating the merit of the proposed model. Thus, this model can be further employed as the fundamental of future research on the optimum graphene orientation in anticorrosion composite coatings.

**Keywords**: Graphene; Orientation; Diffusion; Corrosion; Coating; Modelling

## 1. Introduction

Graphene, after exploration in 2004 [1], has revealed the unrivaled ability to solve a large number of scientific and engineering problems in various fields, such as energy conversion and storage [2-5], medicines [6], water treatment [7], membranes [8], sensors [9], deicing [10], and corrosion [11, 12]. Graphene owes its unique electrical, physical, and chemical properties to its special single-layered ultrathin structure [13]. These distinguished properties of graphene have made it a favorable candidate for anticorrosion coatings [11, 14]. Due to its high impermeability and neutrality to chemical attacks, it has been employed both solely [15] and in combination with other materials for corrosion protection purposes [16]. Many researchers have reported the application of graphene on different metallic surfaces with a satisfactory corrosion protection efficiency [17]. However, some scientists have discovered that monolithic graphene coatings for corrosion protection are not reliable enough because they can be mechanically damaged easily and thereby allow corrosion initiation via defects, losing their excellent anticorrosion performance [18]. To overcome this flaw, researchers have composited graphene within metallic and organic corrosion protective coatings [16, 19]. It has been found that the addition of small quantities of graphene to organic matrices can greatly enhance the anticorrosion performance of the coatings by creating a torturous pathway for corrosive species and some other mechanisms [20].

The unique two-dimensional structure of graphene results in its anisotropic properties [21]. For instance, the electrical and thermal conductivity of graphene are not the same in all directions [22]. Based on this fact, some research groups have investigated the influence of graphene orientation in anticorrosion coatings on their corrosion protection behavior [23, 24]. It has been discovered that graphene sheets in composite coatings create tortuous diffusion paths, thereby making the penetration of corrosive factors in the coatings difficult and leading to augmented resistance to corrosion. Nonetheless, the disordered distribution of graphene is unable to effectively limit corrosion because only graphene sheets parallel to the substrate can provide the absolute shielding effect, whereas perpendicular ones do not fall in the shielding concept. In other words, larger angles between the graphene sheets and underlayer provide less effective protection against corrosion. The engineering of graphene orientation is thus a potential way to further strengthen the protective efficiency of the coatings. The graphene orientation was mainly controlled by the magnetic field orientation of magnetic particles [24], electric field orientation [25], and layer by layer orientation of the self-assembly [26]. For instance, scientists successfully coupled iron oxide nanoparticles, including α-$Fe_2O_3$ and $Fe_3O_4$ with graphene sheets [23, 27]. These magnetic nanoparticles were adsorbed onto the graphene surface by electrostatic interactions and hence magnetized the graphene sheets. The magnetized graphene sheets then became capable of rotating inside epoxy under the influence of a magnetic field and could be aligned in different orientations. It was also concluded that the parallel arrangement of the magnetic graphene sheets greatly enhances the permeation resistance and reduces conductivity, which are beneficial for improving the anticorrosion properties of the coatings [28].

Although the effect of graphene orientation on the "overall" corrosion rate of graphene-containing composite coatings has been previously published, there is no report on distinguishing the "corrosion initiation time" of these coatings, to our knowledge. The corrosion initiation time is referred to the incubation period elapsed before the onset of corrosion reactions and damage at the coating/substrate interface. Knowing the corrosion initiation time under graphene-reinforced composite coatings exposed to corrosive environments warns about the status of the substrate and can be used for inspection and repair schedules. Accordingly, this research work aims to introduce a model to predict the corrosion onset time under graphene-containing composite coatings.

## 2. Model setup

### *2.1. Graphene orientation fundamentals in anticorrosion coatings*

In this model, it is assumed that graphene sheets are all single-layered rigid two-dimensional ones that do not allow any corrosive species to pass through and are impermeable. The diffusion coefficient of graphene in this work is supposed to be zero because it has been previously proved that a monolayer of graphene is impermeable to many materials including standard gases [29]. Moreover, it is supposed that the quantity of graphene in composite coatings is enough to form a single layer in total, covering the whole coated area over the metallic substrate. Graphene in organic coatings can have various orientations. Fig. 1a demonstrates a circumstance in which graphene sheets are all arranged in one layer in the entire organic coating. This layer of graphene is supposed to have a zero angle and to be absolutely impermeable. In Fig. 1b, graphene sheets are aligned at a 45º angle and allow a part of corrosive species to pass through, whereas the graphene sheets with the 90º angle (Fig. 1c) exhibit no resistance against the permeation of infiltrating substances.

In order to ascribe the diffusion of substances to the orientation degree of graphene sheets in composite coatings and derive the corresponding mathematical equations, a geometrical model is proposed (Fig. 1d). Herein, it is supposed that in an element, the lengths of the graphene layer and organic layer are the same ($g$=$L$). By rotating this layer of graphene with the length $g$, only a part of corrosive substances can pass through the organic coating that does not hit the graphene layer. The ratio of the permeating substances is assumed to be proportional to the surface area that is not under the projected surface area of the graphene layer. Thus, in an element, the graphene sheet hinders diffusing substances with a projected layer as long as $g\cos(\theta)$. On the contrary, the organic coating with a surface that does not lie under the graphene layer and is $g(1-\cos(\theta))$ long, allows the corrosive species to permeate freely. Herein, $(1-\cos(\theta))$ is called the unprotected projected surface area proportion factor.

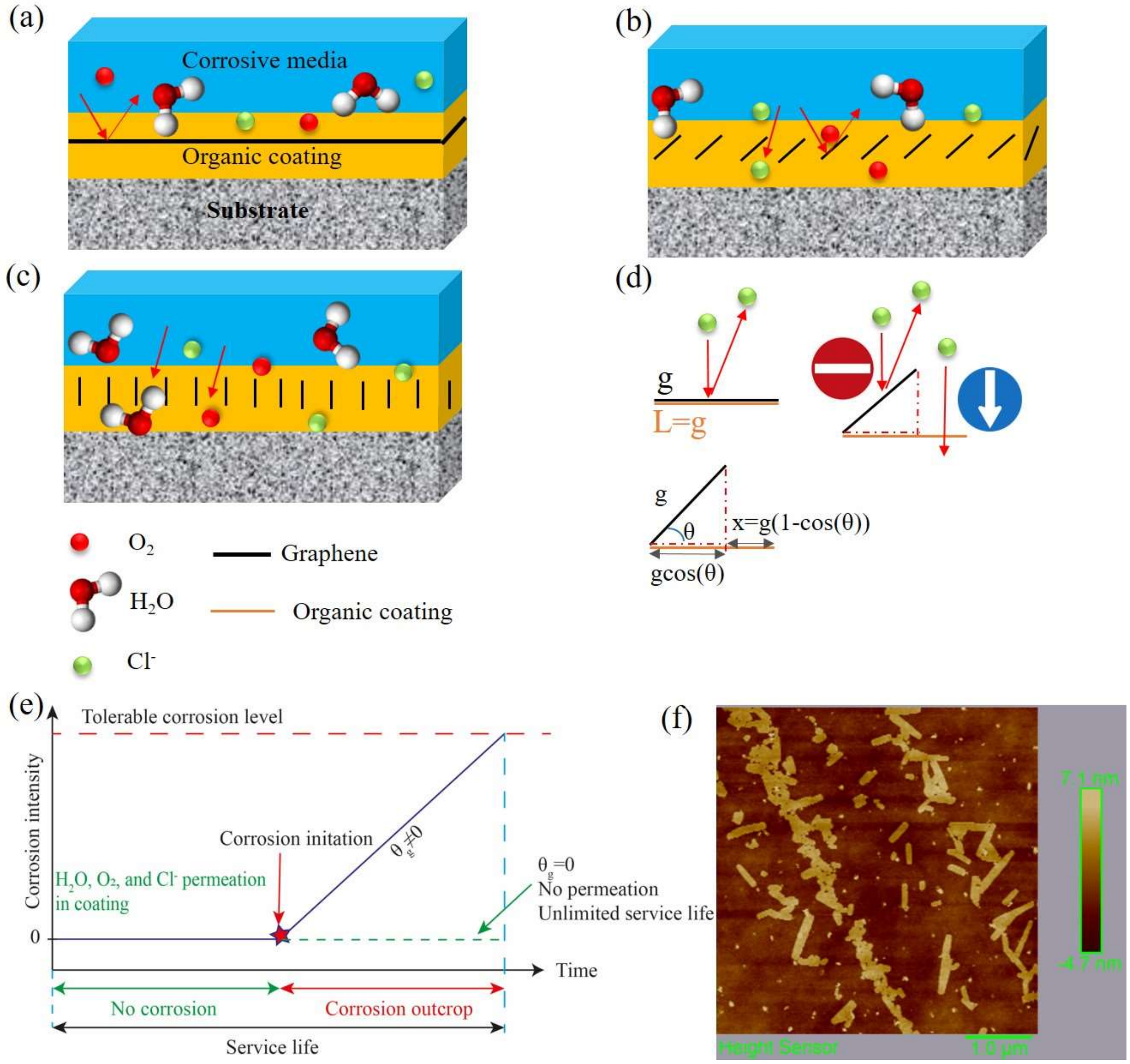

Fig. 1. Schematic illustration of the composite coatings consisting of organic materials and graphene sheets with the graphene orientation of (a) 0, (b) 45, and (c) 90º designed for corrosion prevention; (d) geometry-based diffusion of corrosive species; (e) corrosion evolution stages of the substrates coated with graphene/organic composite coatings (adapted from [30]); (f) AFM image of the purchased graphene sheets dispersed in thinner.

Regarding the above assumptions, the diffusivity of a substance in a composite coating ($D_c$) consisting of an organic material and graphene sheets can be described by Eq. 1.

$$D_c = \frac{A_g}{A_e} D_g + \frac{A_p}{A_e} D_p \quad (1)$$

where $A_g$= the surface area of a graphene sheet, $A_e$= the surface area of the element, $D_g$= the diffusion coefficient of graphene which is zero, $A_p$= the surface area of the polymeric coating, and $D_p$= the diffusion coefficient of the polymeric coating in $cm^2\ s^{-1}$. Taking the effect of geometry into account, as illustrated in Fig. 1d, Eq. 2 is obtained for calculating $D_c$.

$$D_c = (1 - \cos\theta) D_p \quad (2)$$

where $\theta$is the angle between the substrate and graphene sheets.

### *2.2. Flow of corrosive species*

The flux is the rate at which various substances infiltrate inside a material. The flux is actually the number of atoms, ions, molecules, or other particles flowing through a defined area in a specified time. The Fick's first law [31] can describe the net flux of various species inside a composite coating:

$$J = -D_c \frac{\partial C}{\partial x} \quad (3)$$

where $J$ represents the flux in $cm^{-2}\ s^{-1}$ and $\frac{\partial C}{\partial x}$ stands for the concentration gradient in $cm^{-4}$. Substituting Eq. 2 in Eq. 3 for steady-state conditions results in Eq. 4.

$$J = -(1 - \cos\theta) D_p \frac{\Delta C}{\Delta x} \quad (4)$$

where $\frac{\Delta C}{\Delta x}$ is the concentration difference in two points per length.

### *2.3. Diffusion of corrosive species*

In non-steady state conditions, the diffusion of permeating species inside a matrix can be explained by the following differential equation, which is called the Fick's second law [31].

$$\frac{\partial c}{\partial t} = \frac{\partial}{\partial x}\left(D\,\frac{\partial c}{\partial x}\right) \tag{5}$$

where $\frac{\partial c}{\partial t}$ is the concentration variation in the unit of time, *c* is the concentration of the permeating material, and *x* is diffusion distance. If the diffusion coefficient *D*, does not vary with location *x* and $D=D_c$, a simplified version of the Fick's second law can be written as follows [31]:

$$\frac{\partial c}{\partial t} = D_c\frac{\partial^2 c}{\partial x^2} \tag{6}$$

For solving Eq. 6 in realistic conditions, the below three boundary conditions were employed and the solution is exhibited in Eq. 10.

$$C(x,0) = 0 \tag{7}$$

$$C(0,t) = C_s \tag{8}$$

~~$\frac{\partial}{\partial x}C(L,t) = 0$ (9)~~

$$\frac{C_s-C_x}{C_s-C_0} = erf\left(\frac{x}{2\sqrt{D_c t}}\right) \tag{10}$$

where $C_s$ is the fixed concentration of the permeating species at the surface of the probing material, $C_0$ is the uniform concentration of the infiltrating substance at the beginning, and $C_x$ is the concentration of the permeating species at the location *x* in the coating after time *t*. Assuming $C_0 = 0$ and substituting Eq. 2 in Eq. 10, the following formula is derived to determine the concentration of a diffusing substance at location *x* and time *t* with graphene layers rotated at the $\theta$ degree.

$$C_x = C_s\left(1 - erf\left(\frac{x}{2\sqrt{(1-\cos\theta)D_p t}}\right)\right) \tag{11}$$

### *2.4. Corrosion initiation*

As exhibited in Fig. 1e, the corrosion process of materials coated with protective coatings is dividable into two steps: 1) corrosive substances infiltration and 2) corrosion outcrop [30]. The first step includes the penetration of corrosive species, such as water, $Cl^-$, and $O_2$ inside the protective coating, where no damage occurs. The reason for this fact is the slow diffusion process and consequently low concentration of the corrosive substances, which is not high enough for the initiation of corrosion. By enhancing the concentration of the corrosive species at the substrate/coating interface and reaching a threshold, the corrosion phenomenon begins and the substrate starts experiencing corrosion damage. After a period of time, the intensity of corrosion rises to a level that the part needs to be repaired or replaced. The summation of permeation time before reaching a threshold and corrosion outcrop constitutes the service life. On the other hand, when all graphene sheets are aligned horizontally ($\theta_g = 0$), in ideal conditions that no substance can pass through the graphene sheets, corrosion does not take place and the service life will be unlimited.

The concept of the two-step corrosion process is applicable to metals with either active or passive surfaces. The threshold concentration ($C_{th}$) of corrosive species for active materials is any value greater than 0, whereas for passive surfaces this quantity needs to be more than the concentration of the corrosive substance required to deteriorate the passive layer and accelerate corrosion. These two steps can be mathematically defined by Eqs. 12 and 13. During the diffusion stage, the concentration of the corrosive substance at the metal surface ($C(L,t)$) is less than $C_{th}$ from the 0 moment to the time that the concentration reaches the threshold concentration $T_{th}$. After the accumulation of a critical quantity of the corrosive species at the surface of the substrate, corrosion begins at the time $T_{th}$ (Eq. 13) [32].

Diffusion stage $$C(L,t) < C_{th} \; for \; 0 \leq t < T_{th} \quad (12)$$

Corrosion initiation $C(L,t) = C_{th}\ for\ t = T_{th}$ (13)

To determine the time of corrosion onset, corrosion initiation conditions are substituted in Eq. 10 with the same boundary and initial conditions. Therefore, the time required for corrosion initiation can be calculated as follows [32]:

$$T_{th} = f(C_s, C_{th}, D, x, \theta) = \frac{x^2}{4D_c\left[erf^{-1}\left(1-\frac{C_{th}}{C_s}\right)\right]^2} \quad (14)$$

Replacing Eq. 2 in Eq. 14 and substituting $x$ in Eq. 14 with $L$ (the anticorrosion coating thickness), Eq. 15 is achieved to determine the corrosion initiation time as a function of the orientation angle of the graphene sheets in the coating.

$$T_{th} = f(C_s, C_{th}, D, x, \theta) = \frac{L^2}{4D_p(1-\cos\ )\left[erf^{-1}\left(1-\frac{C_{th}}{C_s}\right)\right]^2} \quad (15)$$

## 3. Experimental

### *3.1. Materials*

Pure iron plates (99.9% Fe) were purchased from Dongguan Lingyue Mould Steel Co., Ltd. and used as the substrates. They were polished successively by sand papers ranging from 80 to 10,000 grits, washed with double-distilled water, degreased with acetone, and dried at 90 ºC for 15 min prior to coating. Epoxy resin E-44, isoamyl acetate as the thinner, and glycerol as the hardener were bought from Zhengzhou Penghui Chemical Products Co., Ltd. Graphene powder of 1-3 layers, 7-12 µm in diameter, and 22 $m^2\ g^{-1}$ in specific surface area (Fig. 1f) was also purchased from Xiamen Kaiwei graphene technology Co. Ltd. All reagents were used as received without further purification.

### *3.2. Epoxy-graphene composite coatings fabrication*

70 mg graphene was dispersed in 30 g thinner via sonication for 30 min. Then, the suspension was added to 30 g epoxy resin and sonicated for 1 h. After obtaining a homogenous slurry, the hardener was added to the dispersion at the epoxy/solvent/hardener weight ratio of 1:0.875:0.2, followed by mixing and sonication. Excess solvent was eliminated by placing the suspension in a vacuum chamber for 30 min. Finally, the substrates were coated by dipping into the suspension and subsequently drying at room temperature, 40 and 70 °C, each for 24 h to obtain composite coatings with desired thicknesses (~30-100 μm). Meanwhile, the epoxy coating without graphene was prepared by the same route.

### *3.3. Structure characterization*

In order to study the structure and composition of the coated samples, a Fourier transform infrared spectrometer (FTIR, Bruker, Tensor II, wavenumber = 4000-400 $cm^{-1}$) and X-ray diffractometer (XRD, Shimadzu, XRD-6100, Cu Kα radiation, λ=0.15418 nm, 2θ=10º-90º) were utilized. A field emission scanning electron microscope (FESEM, ZEISS, Sigma 300) was also employed to assess the morphology and graphene dispersion in the matrix. Meantime, the cross-section of the coatings was investigated by FESEM. An atomic force microscope (AFM, Bruker, Dimension Icon) was also employed to assess the morphology of the as-purchased graphene sheets. A porosity analyzer (Micromeritics ASAP 2460) was recruited to determine the Brunauer–Emmett–Teller (BET) specific surface area of the graphene sheets.

### *3.4. Corrosion evaluation*

The corrosion initiation of the coatings in 3.5 wt% NaCl solution was probed on a CHI 660E electrochemical workstation by conducting electrochemical impedance spectroscopy (EIS) at open circuit potential (OCP) from 10000 Hz to 0.01 Hz at 20 mV amplitude. The electrochemical tests were conducted by a conventional three-electrode system consisting of a saturated calomel electrode as the reference electrode, pure iron coated with the various epoxy-matrix deposits as the working electrode, and platinum foil as the counter electrode. ZSimpWin software was also employed for equivalent electrical circuits modelling of the EIS data.

## 4. Results and discussion

### *4.1. Diffusivity and mass flux dependency on the orientation of graphene sheets*

Corrosive substances such as water, oxygen, and chloride ions come from environments like seawater. Graphene sheets in the coating hinder the free infiltration of corrosive substances. As a result, a portion of the corrosive substances can pass through the spaces between the graphene sheets, and the remaining part is hindered by graphene sheets from reaching the protected substrate. For instance, Fig. 2a schematically exhibits the permeation of $Cl^-$ in a graphene/organic composite coating.

The model introduced in this study is based on the diffusion coefficient of corrosive substances as a function of the degree between the graphene sheets and substrate surface (Eq. 2). Hence, this model is able to predict the diffusion coefficient of various substances in different matrices reinforced with a layer of graphene sheets. To obtain the relationship between the diffusion coefficients of the corrosive species and employ that for further calculations, the diffusion coefficient values of chloride, oxygen, and water molecules in two different polymers are extracted from the literature, as tabulated in Table 1.

Table 1. Diffusion coefficient values of three corrosive species in polymers.

| Polymer | Corrosive substance | Diffusion coefficient ($cm^2\ s^{-1}$) | Ref. | Concentration in seawater | Ref. |
|---|---|---|---|---|---|
| E-44 epoxy | $Cl^-$ | $4.68\times10^{-12}$ | [33] | ~21 g $L^{-1}$ | [34] |
| Polystyrene | $O_2$ | $19\times10^{-14}$ | [35] | ~8.5 mg $L^{-1}$ (average value) | [36] |
| Polystyrene | $H_2O$ | $900\times10^{-14}$ | [37] | ~1.025 g $cm^{-3}$ (average value) | [38] |
| E-44 epoxy | $H_2O$ | $2.44\times10^{-9}$ | [33] | | |

Figs. 2b, d, and f display the relationship between the diffusion coefficients of chloride ions in E-44 epoxy and oxygen and water molecules in polystyrene, determined by Eq. 2, respectively. In these polymeric coatings, when there is no graphene or graphene sheets have an orientation degree of 90º, the value of the diffusion coefficient is unchanged because in Eq. 2, the term (1-cos(θ)) equals to one. On the contrary, the zero angle results in no diffusion, which means that no substance can pass through them and results in a diffusion coefficient of 0 $cm^2\ s^{-1}$. The enhancement of θ increases the value of the diffusion coefficient for various substances in the matrix. To be specific, the angles of 30 and 60º decrease the value of the diffusion coefficient to 0.134D and 0.5D, respectively. For instance, the diffusion coefficient of $Cl^-$ at 30º and 60º is predicted to be $0.627\times10^{-12}$ and $2.34\times10^{-12}$ $cm^2\ s^{-1}$, respectively.

Another significant parameter indicating the infiltration process of various substances in a material is mass flux ($J$). The contribution of the graphene sheets angle ($\theta$) in the two different matrices is calculated by Eq. 4, as depicted in Figs. 2c, e, and g for chloride, oxygen, and water molecules, respectively. Since the flux is a function of the aforementioned diffusion coefficient, its absolute value is maximum at θ=90º and minimum (zero) at θ=0º. The more negative quantities of the mass flux indicate larger flux entering the coating. To be specific, the angles of 30 and 60º decrease the value of the mass flux to 0.134J and 0.5J, respectively. For example, the mass flux quantity of $Cl^-$ in E-44 epoxy with graphene sheets oriented at 30 and 60º is predicted to be $1.254 \times 10^{-7}$ and $4.68 \times 10^{-7}$ mg $m^{-2}$ $s^{-1}$, respectively.

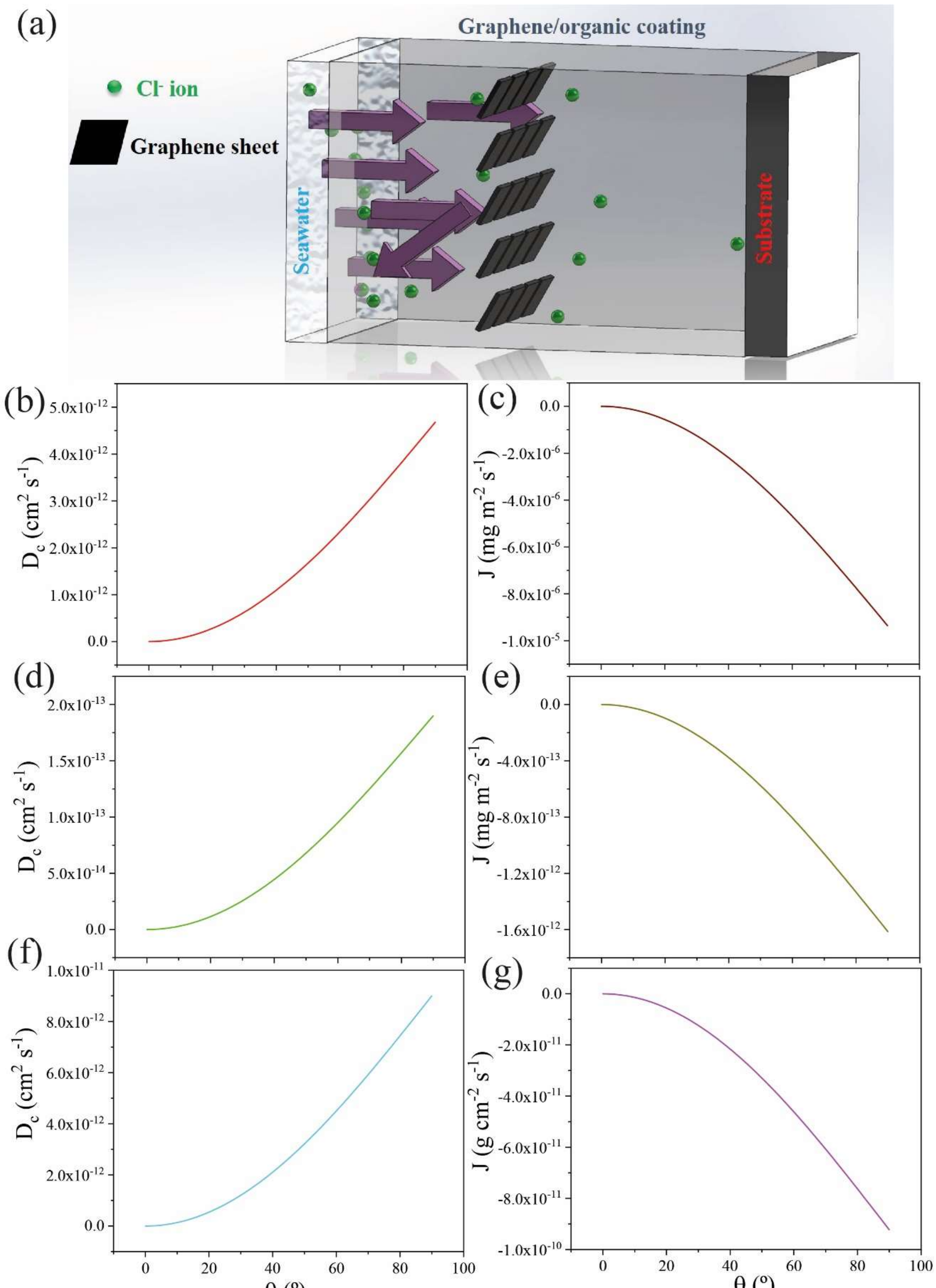
(a)
Graphene/organic coating
Cl⁻ ion
Graphene sheet
Seawater
Substrate
(b)
(c)
(d)
(e)
(f)
(g)
$D_c$ (cm$^2$ s$^{-1}$)
J (mg m$^{-2}$ s$^{-1}$)
J (g cm$^{-2}$ s$^{-1}$)
θ (º)

Fig. 2. (a) 3D schematic illustration of $Cl^-$ ions diffusing into a graphene/organic composite coating; (b) Variation of the diffusion coefficient of chloride in E-44 epoxy as a function of the alignment degree of graphene sheets; (c) Mass flux of $Cl^-$ in E-44 epoxy as a function of the alignment degree of graphene sheets; Variations of the diffusion coefficient of (d) $O_2$ and (f) water molecules in polystyrene vs. the alignment degree of graphene sheets; the mass flux of (e) $O_2$ and (g) water molecules in polystyrene vs. the alignment degree of graphene sheets.

### *4.2. Concentration profile vs. graphene orientation*

Figs. 3a, b, and c display the one-dimensional concentration profiles of chloride ions inside a graphene/E-44 epoxy coating, those of oxygen and water molecules inside a graphene/polystyrene coating at different angles. The plotted figures are surfaces standing for the concentration of the permeating species as a function of time and distance from the surface (Eq. 11). To plot these concentration profiles, Matlab software was used. In order to conduct the calculations using ordinary computers, the diffusion coefficients were amplified, and the time periods were minified by the same ratio. The real time scales employed in the computations are shown in each figure.

According to the model introduced in this study, at 0º the concentration of all corrosive substances regardless of their physical properties inside the whole coating is zero because no permeation is predicted at this angle. Despite the excellent performance of graphene sheets with the zero angle, the sheets with $\theta_g$=90º (parallel to the diffusion direction) have no influence on diffusion, and corrosive substances can reach the substrate without experiencing any resistance from the graphene sheets in the coating. The decline of the orientation degree of graphene sheets to 60º effectively reduces the concentration level of the probing diffusive substances in the composite coatings. The more reduction of $\theta_g$ to 30º lowers the level of the concentration surface greatly to a very low level in a way that the concentration of the corrosive substances beyond 0.5 cm during the investigated time periods would be zero.

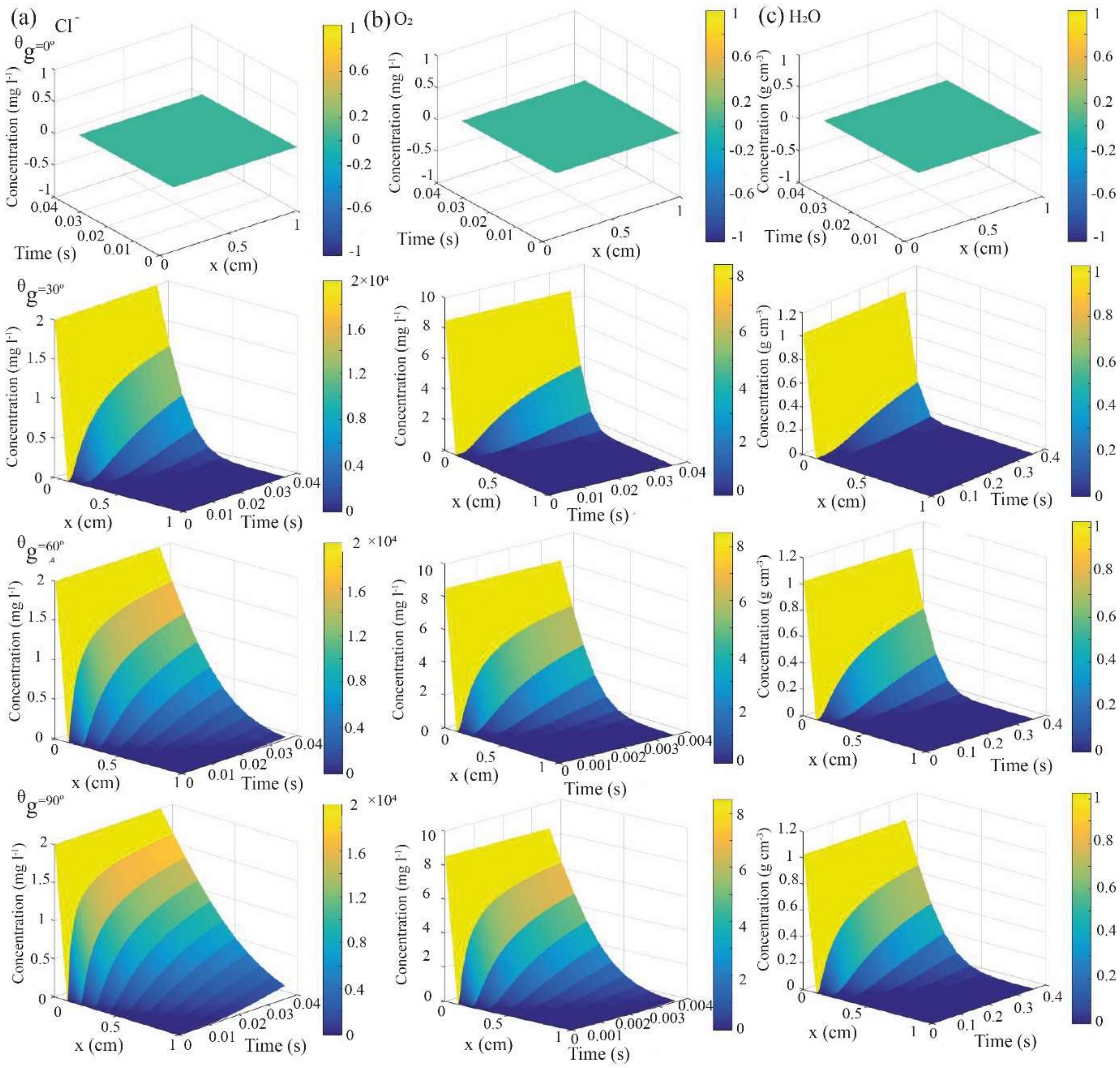

Fig. 3. Concentration profile of (a) $Cl^-$ in the graphene/E-44 epoxy composite, (b) $O_2$ in the graphene/polystyrene composite, and (c) $H_2O$ in the graphene/polystyrene composite, calculated by the Fick's second law for the different orientation angles of graphene sheets.

### *4.3. Coating characterization*

To evaluate the applicability of the model introduced in this study, E-44 epoxy and its composite with graphene were coated on the pure iron samples and tested. The experimental data was then compared with the predicted values. SEM images (Fig. 4a and Fig. 4d) exhibit crack-free coatings of the epoxy/graphene composite and epoxy, respectively. These coatings were well attached to the substrate without any gap and were approximately 66 and 81 μm thick (Fig. 4b and Fig. 4e), respectively. Graphene sheets in the composite are also observable in Fig. 4c.

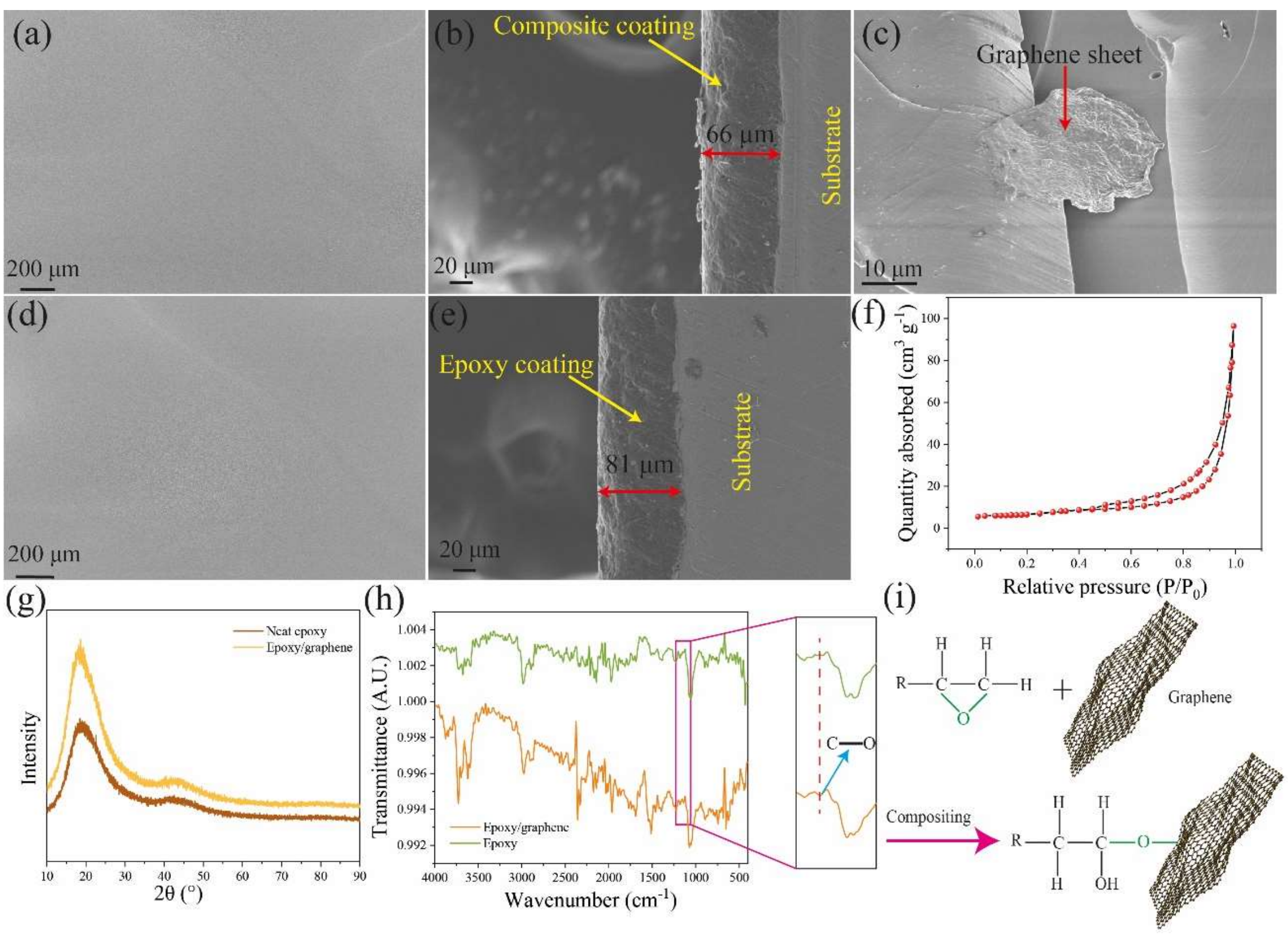

Fig. 4. SEM images of the graphene/E-44 epoxy composite: (a) surface, (b) cross-section, and (c) graphene sheets in the matrix; SEM images of the (d) surface and (e) cross-section of the epoxy-coated metallic samples (The SEM images were captured using secondary electron

mode at the working voltage of 10 kV.); (f) BET nitrogen adsorption–desorption isotherm; (g) XRD patterns and (h) FTIR spectra of the neat E-44 coating and the composite coating; and (i) Schematic illustration of graphene bonds with the E-44 epoxy resin.

The BET nitrogen adsorption–desorption isotherm of the graphene powder prior to dispersion in the thinner is also exhibited in Fig. 4f. The powder presented a BET specific surface area of 22 $m^2g^{-1}$. The nitrogen adsorption–desorption isotherm belongs to type V isotherms which are characterized as pore condensation and hysteresis. Furthermore, the initial part of this sorption isotherm is ascribed to the adsorption isotherm of type III, suggesting the absence of micropores.

Fig. 4g exhibits the XRD patterns of the epoxy/graphene nanocomposite and neat epoxy. The sharp peak-free XRD pattern of the neat epoxy reveals its amorphous structure, where the addition of graphene did not alter its XRD pattern. Typically, the lack of graphene characteristics peaks in the composite coatings reveals the uniform dispersion of graphene sheets in the matrix due to its low content. Fig. 4h also displays the FTIR spectra of the neat epoxy and epoxy/graphene composite. The FTIR pattern of the composite reveals an additional absorption peak at 1140 $cm^{-1}$ as compared with the neat epoxy. This peak is ascribed to the presence of the C−O stretching vibration mode of the aliphatic ether [39, 40], revealing the existence of aliphatic ethers in the composite (Fig. 4i). The polyamine in the curing agent catalyzes the reaction of hydroxyl groups on graphene sheets with the epoxy group of the epoxy monomer during the in-situ polymerization process. Moreover, the intensity of C=O stretch of the COOH group at 1691 $cm^{-1}$ and carboxylic functional group at 3412 and 3498 $cm^{-1}$ is strengthened by the addition of graphene sheets to the epoxy matrix, which is compatible with Ref. [41].

### *4.4. Impedance spectroscopy*

The use of equivalent electrical circuits (EECs) for investigating corrosion mechanisms is a credible and universal technique. Researchers propose an appropriate EEC for their systems regarding the shape of impedance diagrams (Nyquist, Bode-impedance magnitude, and Bode-phase angle plots) and the instinct of the surface. Moreover, any obvious alteration in the form of such diagrams is a sign of changes in their corresponding EEC.

To improve fittings in the EEC modelling of the impedance data in this study, constant phase element (CPE) was utilized rather than "ideal" capacitors, with an impedance depending on frequency:

$$\frac{1}{Z_{CPE}} = Q(j\omega)^n \qquad (16)$$

where $Z_{\mathrm{CPE}}$ means impedance, $Q$ and $\omega$ show pseudo-capacitance and angular frequency, respectively, and $n$ is the power coefficient related to homogeneity ($0 < n \leq 1$, 1 is for the ideal capacitance).

The Nyquist and bode EIS results of the samples and their corresponding EECs are presented in Fig. 5 a-f and g, respectively. Routinely, immediately after immersing the coated samples inside the electrolyte, the organic coatings exhibit a behavior predicted by the Randles circuit (Fig. 5 g). This circuit is composed of the solution resistance, $R_s$ in series with the coating or pore resistance, $R_c$, and coating capacitance, $C_c$. However, the rapid diffusion of corrosive species in the coatings challenges the ideal Randles circuit. Thus, the EIS curves of the epoxy-based coatings obtained in this study after several hours of immersion in the electrolyte were fit much better with the Model 1 circuit in comparison with the typical Randles circuit. This model affirms the onset of attack to the substrate with the penetration of the electrolyte. For instance, in a 100 μm E-44 coating, water can reach the substrate in 50 min and initiate slight corrosion [33].

The deviations from the ideal shape of the impedance curve of the Randles circuit in Figs. 5a (epoxy coating of 30 μm in thickness) and 5d (epoxy/graphene coating of 50 μm in thickness) reveal that water molecules had reached the metal/coating interface of the samples within 1 h and 2 days, respectively. This deviation from the ideal conditions is perceivable from the emergence of a small tail in the low-frequency region of the Nyquist plots. The corrosion at this time point happened only in the presence of iron, water, and oxygen molecules, where the $Cl^-$ concentration was below its threshold value for corrosion initiation by $Cl^-$. This is ascribed to diffusion processes caused by the precipitated corrosion products on the substrate/coating interface (represented by W in the EEC) on the surface of active electrochemical sites under the coating. The simulation result recruiting Model EEC 1 exhibited in Fig. 5a affirms a suitable fitting when Model 1 was applied for this stage.

By progression of the exposure of the E-44 and epoxy/graphene samples to the aqueous NaCl solution, the single semicircle shape of their Nyquist plots is replaced by obvious double semicircles and the capacitive arc was dramatically reduced in size after 20 h and 7 days, respectively. Therefore, the corresponding EEC was altered from Model 1 to Model 2, as used for other organic coating [42, 43]. It is noteworthy that these two coatings had different thicknesses, which results in different Nyquist profile alteration time points. $R_{ct}$ and $C_{dl}$ in Model 1 reveal the charge transfer resistance of electrochemical corrosion and double-layer capacitance, respectively. The EEC of the samples damaged by $Cl^-$ (Model 2) in addition to $R_s$, $C_c$, and $R_c$, also includes the resistance of the $Cl^-$-permeated outer film (salt film, $R_{sf}$) in parallel to the salt film capacitance ($C_{sf}$), in series with the resistance of the inner oxide film ($R_{of}$) in parallel to the inner oxide film capacitance ($C_{of}$). $R_{sf}$ is affirmable from the high-frequency loop in Fig. 5b, d, and e. Bessone et al. [44] reported that the high-frequency loop in Nyquist plots are hardly observable in chloride-free solutions. As a result, $R_{sf}$ and the first loop are ascribed to the contribution of $Cl^{-}$ ion. Lee and Pyun [42, 43] also pointed out that the presence of this loop reveals the formation of a chloride-incorporated layer at the surface of metallic substrate during the incubation time of pitting.

Moreover, Figs. 5c and 5f show that the |Z| value in the Bode diagram is sharply decreased by a few orders of magnitude for both samples. These phenomena are ascribed to the aggressive chemical/electrochemical attack of $Cl^{-}$ to the iron substrate after reaching the threshold concentration and consumption of the accumulated formerly existing corrosion products on the substrate surface with an accelerated rate. Thus, it can be concluded that at this stage, chloride ions attack the substrate and deteriorate the coating's corrosion protection property intensively by reaching the threshold level.

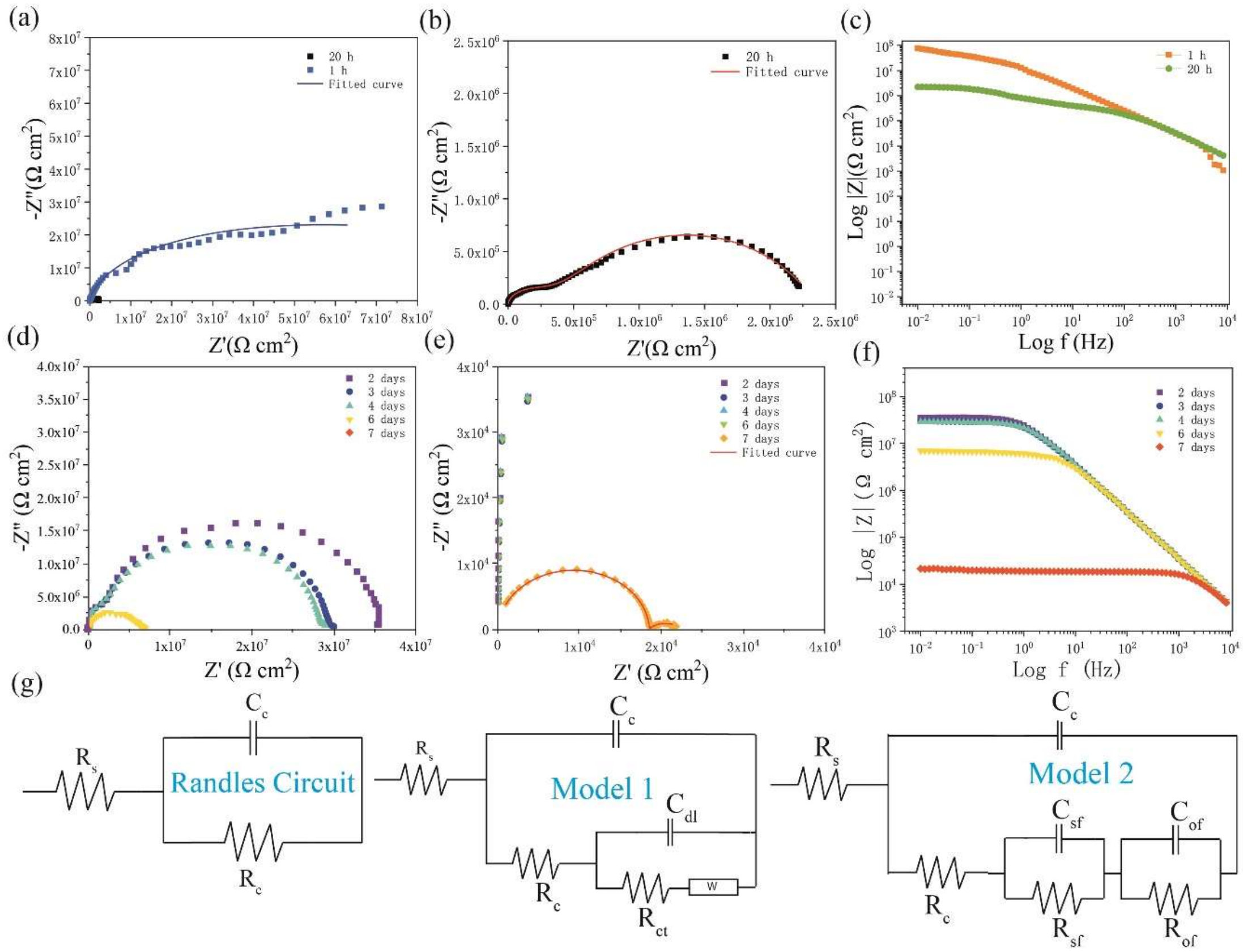

Fig. 5. (a) and (b) Nyquist and (c) Bode-impedance plots of the epoxy-coated iron samples after various immersion times in the electrolyte; (d) and (e) Nyquist and (f) Bode-impedance plots of the epoxy/graphene composite-coated iron samples after various immersion times in the electrolyte; and (g) EECs used to fit the experimental data.

### *4.5. Model predication regarding corrosion occurrence*

The diffusion of corrosive substances inside anticorrosive coatings deposited on engineering materials is a determining factor in the onset of the corrosion process and subsequent failure. For instance, the corrosion of carbon steels in seawater often starts with the uptake of water and oxygen molecules and is accelerated dramatically by the attack of $Cl^-$. Chloride ions dissolved in seawater permeate into anticorrosive coatings and after a period of time, reach the surface of the steel. It is noteworthy that owing to the lower diffusion coefficient of $Cl^-$ in typical organic coatings in comparison to water, corrosion reactions in the metal/coating interface in the presence of this ion occur after corrosion caused by water molecules. The introduction of $Cl^-$ beyond the threshold level (500 ppm on the surface of steel carbon) is able to accelerate the corrosion attack [45].

Since the orientation angle of graphene sheets greatly influences the diffusion, flux, and concentration distribution of corrosive species, the time when the $Cl^-$ concentration is high enough to start the corrosion process can be determined. Eq. 15 displays the mathematical relationship between the corrosion threshold time, $T_{th}$, and the orientation degree of the graphene sheets in the composite coatings. The high and low limits of the critical concentration of $Cl^-$ were obtained from the literature ($10^{-4}$ M NaCl-$3\times10^{-3}$ M NaCl) [46-48]. Fig. 6a exhibits the logarithmic relationship between the corrosion initiation time ($T_{th}$) of a pure iron part coated with 100 μm thick graphene/E-44 epoxy anticorrosion coating in seawater and the orientation angle of the graphene sheets. As in the seawater $Cl^-$ is known to be the most harmful corrosive substance, the parameters of Eq. 15 are replaced with those of $Cl^-$ in seawater. According to our predictions, the orientation angle of 90º does not have any influence on $T_{th}$, whereas the smaller angles enhance $T_{th}$. At 90º, the corrosion onset time is predicted to be 214.5 h for the lower critical concentration of $Cl^-$. The orientation of 45 º, which is indicative of the random orientation of the graphene sheets, can postpone the corrosion initiation by about 3.4 times. The corrosion onset time with a smaller angle of 10º is predicted to be approximately 65.8 times larger than that of the parallel alignment with the diffusion direction. As $\theta_g$ tends to zero, $T_{th}$ approaches infinite time, which means that at this angle under ideal conditions, the corrosive substances will never reach the substrate; hence, corrosion will not occur at all.

## 5. Model verification

To evaluate the validity of the corrosion initiation time and the effect of the graphene orientation on the corrosion initiation predicted by the proposed model, the data obtained from the experiments are compared with the calculated values. The corrosion initiation time by $Cl^-$ attack to the substrate coating is obtained from the Nyquist plots, when the regime varies significantly as indicative of the accumulation of $Cl^-$ above its critical concentration.

Figs. 6b and 6c exhibit the corrosion onset vs. the orientation degree of graphene sheets in 30 and 50 µm thick composite coatings, respectively. The model estimates that the corrosion of iron under the 30 µm thick epoxy coating starts at 18 h to 33 h of exposure (Fig. 6b). It is noteworthy that the maximum and minimum $T_{th}$ values were calculated by substituting the two highest and lowest values of $C_{th}$ obtained from the literature in Eq. 15. The tested sample showed obvious corrosion onset by $Cl^-$ invasion after 20 h. The corrosion onset of the sample coated by the epoxy/graphene composite was also expected to begin at 180 h to 321 h (supposing 45° for the orientation degree of graphene sheets in the epoxy matrix). The tested value was 168 h, which was very close to the model estimation (Fig. 6c).

The optimal corrosion protection found for the aligned arrangement of graphene sheets in some experimental studies [25, 49] also verifies the prediction of the model introduced in this paper. Typically, Ding et al. [49] discovered that the horizontal arrangement of magnetic graphene significantly enhances the diffusion resistance of corrosive substances and hence improves the corrosion resistance of composite coatings. In addition, the relationship established between the alignment degree of the graphene sheets and corrosion prevention motivates scientists to invent and employ new technologies toward setting the orientation of graphene sheets in the incorporated coatings, aiming at using the best of graphene and other 2D materials configuration in anticorrosion composite coatings.

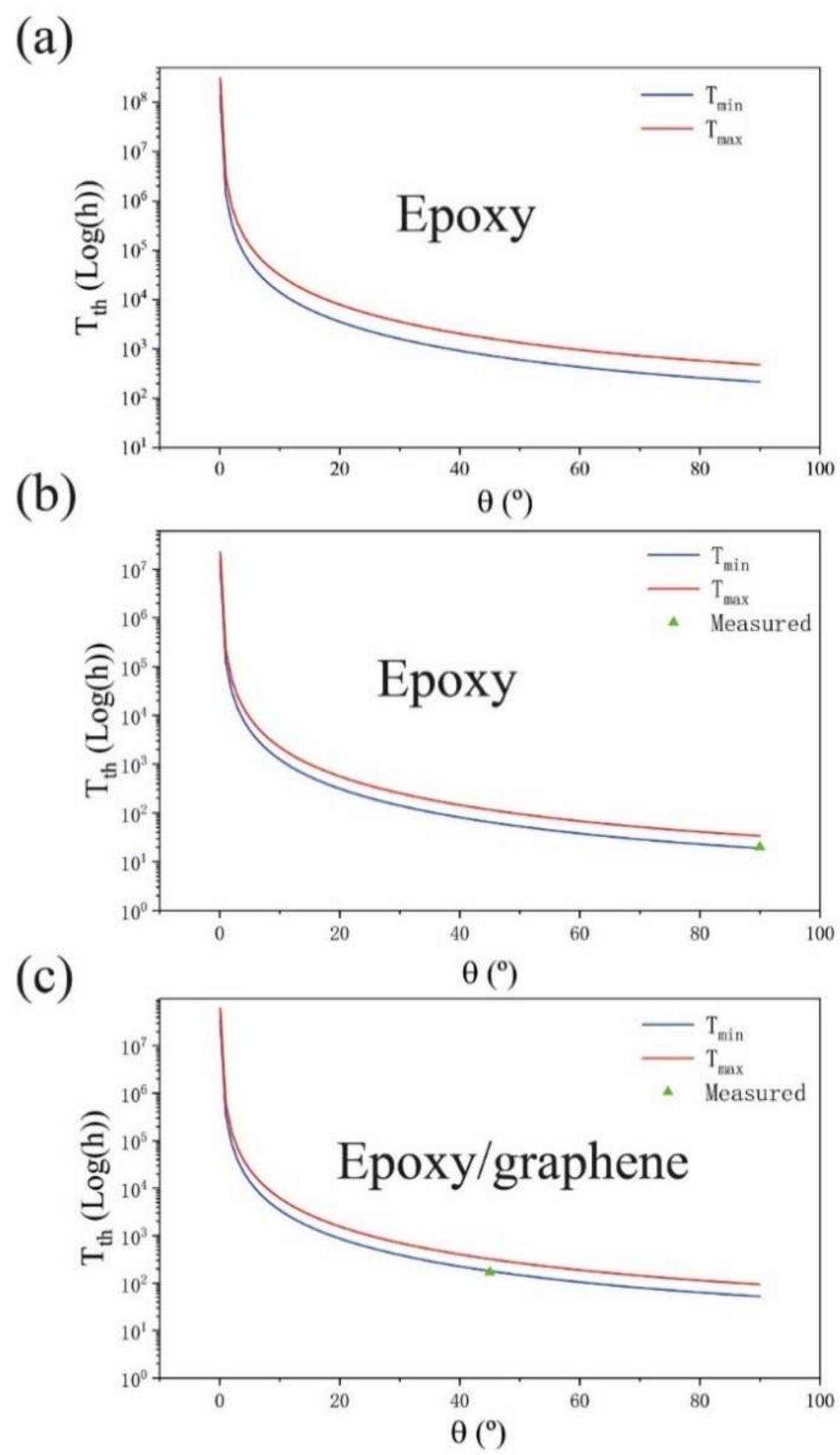

Fig. 6. Corrosion initiation time prediction of iron coated with the graphene/E-44 epoxy composite coating with (a) 100 μm, (b) 30 μm, and (c) 50 μm thickness exposed to the 3.5 wt.% NaCl solution vs. the orientation angle of the graphene sheets.

## 6. Conclusions

The current research introduced a novel model taking the effect of the graphene orientation in anticorrosion composite coatings into account to estimate the diffusion and flux of corrosive substances in the coating as well as the corresponding corrosion initiation time and service life. This model was developed based on the principal Fick's first and second laws related to the diffusion phenomena, while simultaneously considering the graphene sheets orientation through a novel trigonometric factor called "unprotected projected surface area proportion". The model revealed that the reduction of the orientation degree between graphene sheets in composite coatings and the substrate slows down the diffusion process and flux. As a result, the corrosion is estimated to initiate after longer durations, allowing the underlayer to enjoy a prolonged service life. It was estimated that the 10º angle decelerates the diffusion process significantly and enhances the service life approximately 65 times in comparison with the 90º alignment for a 100 μm thick epoxy/graphene composite coating. Moreover, the parallel alignment of graphene sheets in a matrix under ideal conditions can absolutely hinder the diffusion, flux, and corrosion, thereby leading to an unlimited service life of the protected part. The validity of the predicted quantities was also confirmed by experimental data in terms of closeness. Therefore, the relationships and values predicted in this work have the potential to be the basis of further experimental and theoretical developments towards finding the optimum graphene orientation degree in anticorrosion composite coatings.

## Acknowledgement

The authors would like to thank Chongqing 2D Materials Institute for its generous support.

## References

[1] K.S. Novoselov, A.K. Geim, S.V. Morozov, D. Jiang, Y. Zhang, S.V. Dubonos, I.V. Grigorieva, A.A. Firsov, Electric Field Effect in Atomically Thin Carbon Films, Science 306(5696) (2004) 666.
[2] M. Shaker, R. Riahifar, Y. Li, A review on the superb contribution of carbon and graphene quantum dots to electrochemical capacitors' performance: Synthesis and application, FlatChem 22 (2020) 100171.
[3] A.A. Sadeghi Ghazvini, E. Taheri-Nassaj, B. Raissi, R. Riahifar, M. Sahba Yaghmaee, M. Shaker, Co-electrophoretic deposition of Co3O4 and graphene nanoplates for supercapacitor electrode, Materials Letters 285 (2021) 129195.

[4] M. Shaker, A.A.S. Ghazvini, M.S. Yaghmaee, R. Riahifar, B. Raissi, W. Cao, Q. Ge, B. Wang, Prediction of size- and shape-dependent lithium storage capacity of carbon nano-spheres (quantum dots), Journal of Nanoparticle Research 23(8) (2021) 176.
[5] M. Shaker, A.A.S. Ghazvini, W. Cao, R. Riahifar, Q. Ge, Biomass-derived porous carbons as supercapacitor electrodes – A review, New Carbon Materials 36(3) (2021) 546-572.
[6] S. Gooneh-Farahani, M.R. Naimi-Jamal, S.M. Naghib, Stimuli-responsive graphene-incorporated multifunctional chitosan for drug delivery applications: a review, Expert opinion on drug delivery 16(1) (2019) 79-99.
[7] H. Wang, X. Mi, Y. Li, S. Zhan, 3D graphene - based macrostructures for water treatment, Advanced Materials 32(3) (2020) 1806843.
[8] X. Leng, S. Chen, K. Yang, M. Chen, M. Shaker, E.E. Vdovin, K.S. Novoselov, D.V. Andreeva, Technology and Applications of Graphene Oxide Membranes, Surface Review and Letters (2020) 2140004.
[9] S. Wei, X. Qiu, J. An, Z. Chen, X. Zhang, Highly sensitive, flexible, green synthesized graphene/biomass aerogels for pressure sensing application, Composites Science and Technology 207 (2021) 108730.
[10] P. Wang, T. Yao, Z. Li, W. Wei, Q. Xie, W. Duan, H. Han, A superhydrophobic/electrothermal synergistically anti-icing strategy based on graphene composite, Composites Science and Technology 198 (2020) 108307.
[11] X. Leng, S. Chen, K. Yang, M. Chen, M. Shaker, E.E. Vdovin, K.S. Novoselov, D.V. Andreeva, Introduction to Two-Dimensional Materials, Surface Review and Letters (2021) 2140005.
[12] V.Z. Asl, S.F. Chini, J. Zhao, Y. Palizdar, M. Shaker, A. Sadeghi, Corrosion properties and surface free energy of the Zn-Al LDH/rGO coating on MAO pretreated AZ31 magnesium alloy, Surface and Coatings Technology (2021) 127764.
[13] K.S. Novoselov, D.V. Andreeva, W. Ren, G. Shan, Graphene and other two-dimensional materials, Frontiers of Physics 14(1) (2019) 13301.
[14] L. Cheng, C. Liu, D. Han, S. Ma, W. Guo, H. Cai, X. Wang, Effect of graphene on corrosion resistance of waterborne inorganic zinc-rich coatings, Journal of Alloys and Compounds 774 (2019) 255-264.
[15] Y. Wu, W. Zhao, X. Zhu, Q. Xue, Improving the corrosion resistance of graphene-coated copper via accurate defect healing without sacrificing electronic conductivity, Carbon 153 (2019) 95-99.
[16] A. Toosinezhad, M. Alinezhadfar, S. Mahdavi, Cobalt/graphene electrodeposits: Characteristics, tribological behavior, and corrosion properties, Surface and Coatings Technology 385 (2020) 125418.
[17] G. Cui, Z. Bi, R. Zhang, J. Liu, X. Yu, Z. Li, A comprehensive review on graphene-based anti-corrosive coatings, Chemical Engineering Journal 373 (2019) 104-121.
[18] L. Camilli, F. Yu, A. Cassidy, L. Hornekær, P. Bøggild, Challenges for continuous graphene as a corrosion barrier, 2D Materials 6(2) (2019) 022002.
[19] S. Liu, T. Pan, R. Wang, Y. Yue, J. Shen, Anti-corrosion and conductivity of the electrodeposited graphene/polypyrrole composite coating for metallic bipolar plates, Progress in Organic Coatings 136 (2019) 105237.
[20] Y. Ye, H. Chen, Y. Zou, Y. Ye, H. Zhao, Corrosion protective mechanism of smart graphene-based self-healing coating on carbon steel, Corrosion Science (2020) 108825.
[21] E. Pop, V. Varshney, A.K. Roy, Thermal properties of graphene: Fundamentals and applications, MRS Bulletin 37(12) (2012) 1273-1281.
[22] J. Nilsson, A.C. Neto, F. Guinea, N. Peres, Electronic properties of graphene multilayers, Physical Review Letters 97(26) (2006) 266801.

[23] X. Yang, X. Zhang, Y. Ma, Y. Huang, Y. Wang, Y. Chen, Superparamagnetic graphene oxide–Fe3O4nanoparticles hybrid for controlled targeted drug carriers, Journal of Materials Chemistry 19(18) (2009) 2710-2714.
[24] J. Zhou, Q. Wang, Q. Sun, X.S. Chen, Y. Kawazoe, P. Jena, Ferromagnetism in Semihydrogenated Graphene Sheet, Nano Letters 9(11) (2009) 3867-3870.
[25] H. Pang, C. Chen, Y.-C. Zhang, P.-G. Ren, D.-X. Yan, Z.-M. Li, The effect of electric field, annealing temperature and filler loading on the percolation threshold of polystyrene containing carbon nanotubes and graphene nanosheets, Carbon 49(6) (2011) 1980-1988.
[26] L. Zhi, J. Wu, J. Li, U. Kolb, K. Müllen, Carbonization of Disclike Molecules in Porous Alumina Membranes: Toward Carbon Nanotubes with Controlled Graphene ‐ Layer Orientation, Angewandte Chemie 117(14) (2005) 2158-2161.
[27] H. Yan, R. Wang, Y. Li, W. Long, Thermal Conductivity of Magnetically Aligned Graphene–Polymer Composites with Fe3O4-Decorated Graphene Nanosheets, Journal of Electronic Materials 44(2) (2015) 658-666.
[28] J. Renteria, S. Legedza, R. Salgado, M.P. Balandin, S. Ramirez, M. Saadah, F. Kargar, A.A. Balandin, Magnetically-functionalized self-aligning graphene fillers for high-efficiency thermal management applications, Materials & Design 88 (2015) 214-221.
[29] J.S. Bunch, S.S. Verbridge, J.S. Alden, A.M. van der Zande, J.M. Parpia, H.G. Craighead, P.L. McEuen, Impermeable Atomic Membranes from Graphene Sheets, Nano Letters 8(8) (2008) 2458-2462.
[30] K. Tuutti, Corrosion of steel in concrete, Cement-och betonginst.1982.
[31] D.R. Askeland, P.P. Fulay, Essentials of Materials Science & Engineering (SI Version), Cengage Learning2009.
[32] J. Zhang, Z. Lounis, Sensitivity analysis of simplified diffusion-based corrosion initiation model of concrete structures exposed to chlorides, Cement and Concrete Research 36(7) (2006) 1312-1323.
[33] J. Hu, J. Zhang, C. Cao, Determination of water uptake and diffusion of Cl− ion in epoxy primer on aluminum alloys in NaCl solution by electrochemical impedance spectroscopy, Progress in Organic Coatings 46(4) (2003) 273-279.
[34] K.C. Kang, P. Linga, K.-n. Park, S.-J. Choi, J.D. Lee, Seawater desalination by gas hydrate process and removal characteristics of dissolved ions (Na+, K+, Mg2+, Ca2+, B3+, Cl−, SO42−), Desalination 353 (2014) 84-90.
[35] J. Brandrup, E. Immergut, E. Grulke, Polymer Handbook, Wiley, New York, 1999.
[36] R. Baboian, Corrosion Tests and Standards: Application and Interpretation, ASTM International2005.
[37] N.S. Sangaj, V.C. Malshe, Permeability of polymers in protective organic coatings, Progress in Organic Coatings 50(1) (2004) 28-39.
[38] T.S. Garrison, Essentials of Oceanography, Cengage Learning2012.
[39] S. Stankovich, R.D. Piner, S.T. Nguyen, R.S. Ruoff, Synthesis and exfoliation of isocyanate-treated graphene oxide nanoplatelets, Carbon 44(15) (2006) 3342-3347.
[40] W. Zheng, W. Chen, S. Ren, Y. Fu, Interfacial structures and mechanisms for strengthening and enhanced conductivity of graphene/epoxy nanocomposites, Polymer 163 (2019) 171-177.
[41] M. Naebe, J. Wang, A. Amini, H. Khayyam, N. Hameed, L.H. Li, Y. Chen, B. Fox, Mechanical property and structure of covalent functionalised graphene/epoxy nanocomposites, Scientific Reports 4 (2014) 4375.
[42] W.-J. Lee, S.-I. Pyun, Effects of hydroxide ion addition on anodic dissolution of pure aluminium in chloride ion-containing solution, Electrochimica Acta 44(23) (1999) 4041-4049.
[43] W.-J. Lee, S.-I. Pyun, Effects of sulphate ion additives on the pitting corrosion of pure aluminium in 0.01 M NaCl solution, Electrochimica Acta 45(12) (2000) 1901-1910.

[44] J. Bessone, C. Mayer, K. Jüttner, W.J. Lorenz, AC-impedance measurements on aluminium barrier type oxide films, Electrochimica Acta 28(2) (1983) 171-175.
[45] D. Trejo, R.G. Pillai, Accelerated chloride threshold testing: part I-ASTM A 615 and A 706 reinforcement, Materials Journal 100(6) (2003) 519-527.
[46] H.H. Strehblow, Nucleation and repassivation of corrosion pits for pitting on iron and nickel, Materials and Corrosion 27(11) (1976) 792-799.
[47] J.R. Galvele, Transport processes and the mechanism of pitting of metals, Journal of the Electrochemical Society 123(4) (1976) 464.
[48] V. Soulié, F. Lequien, F. Ferreira ‐ Gomes, G. Moine, D. Feron, P. Prene, H. Moehwald, T. Zemb, H. Riegler, Salt ‐ induced iron corrosion under evaporating sessile droplets of aqueous sodium chloride solutions, Materials and Corrosion 68(9) (2017) 927-934.
[49] R. Ding, S. Chen, N. Zhou, Y. Zheng, B.-j. Li, T.-j. Gui, X. Wang, W.-h. Li, H.-b. Yu, H.-w. Tian, The diffusion-dynamical and electrochemical effect mechanism of oriented magnetic graphene on zinc-rich coatings and the electrodynamics and quantum mechanics mechanism of electron conduction in graphene zinc-rich coatings, Journal of Alloys and Compounds 784 (2019) 756-768.